\documentclass[twocolumn,amsmath,trackchanges]{aastex702}

\usepackage{amssymb} 

\usepackage{amsmath}

\begin{document}

\title{Systematic Uncertainties in Rocky Exoplanet Reflected Light from Incomplete Laboratory BRDF Measurements}

\correspondingauthor{Leonardos Gkouvelis}
\email{gkouvelis@iaa.es}

\author[0000-0002-1397-8169]{Leonardos Gkouvelis}
\affiliation{Instituto de Astrofísica de Andalucía (IAA-CSIC), Glorieta de la Astronomía s/n, E-18008 Granada, Spain}
\email{}

\author[0000-0002-8441-2488]{Michael K. Shepard}
\affiliation{Department of Physical and Environmental Sciences, Commonwealth University of Pennsylvania, Hartline Science Center, Room 126, Bloomsburg, PA 17815, USA}
\email{}

\author[0000-0003-4870-1300]{Paul Helfenstein}
\affiliation{Cornell Center for Astrophysics and Planetary Science, Cornell University, Ithaca, NY 14853-6801, USA}
\email{}
\author[0000-0003-1572-7707]{Francisco J. Pozuelos}
\affiliation{Instituto de Astrofísica de Andalucía (IAA-CSIC), Glorieta de la Astronomía s/n, E-18008 Granada, Spain}
\email{}

\author[0000-0003-2215-8485]{Renyu Hu}
\affiliation{Department of Astronomy \& Astrophysics, The Pennsylvania State University, University Park, PA 16802, USA}
\affiliation{Center for Exoplanets and Habitable Worlds, The Pennsylvania State University, University Park, PA 16802, USA}
\affiliation{Institute for Computational and Data Science, The Pennsylvania State University, University Park, PA 16802, USA}
\affiliation{Jet Propulsion Laboratory, California Institute of Technology, Pasadena, CA 91109, USA}
\email{}

\begin{abstract}

The first observations of short-period rocky exoplanets have initiated the era of terrestrial exoplanet characterization. The interpretation of these observations relies on forward models used in atmospheric and surface retrieval frameworks. While atmospheric modeling is supported by extensive laboratory datasets of molecular opacities, the experimental basis for rocky surface characterization remains limited. Most reflectance libraries of planetary analog materials provide spectra measured at a single illumination and viewing geometry, constraining the spectral behavior but not its full angular scattering properties. We show that incomplete angular coverage in laboratory bidirectional reflectance measurements introduces an intrinsic systematic uncertainty in the interpretation of rocky exoplanet observations. Using the Hapke photometric framework, we demonstrate that multiple combinations of photometric parameters can reproduce the same laboratory reflectance spectrum while predicting different angular scattering behaviors and disk-integrated observables. We propagate this non-uniqueness to reflected-light phase curves, geometric albedos, and planet-star contrast ratios between $0.5$--$1.7\,\mu$m, representative of the spectral range targeted by future direct-imaging missions such as the Habitable Worlds Observatory. We find that single-geometry laboratory spectra produce large uncertainties in predicted signals, particularly at small phase angles where opposition effects dominate. Simplified assumptions commonly adopted in exoplanet studies, such as isotropic scattering or neglecting opposition effects, can additionally introduce large systematic biases. Using multi-angular laboratory measurements from the Bloomsburg University Goniometer dataset, we show that the uncertainty in predicted observables collapses once angular scattering behavior is directly constrained. Our results demonstrate that multi-angular measurements are required to construct physically predictive reflected-light models of rocky exoplanets.

\end{abstract}


\section{Introduction}
\label{sec:intro}

Observations of short-period rocky exoplanets with radii $\lesssim 1.6\,R_{\oplus}$ have recently become possible with the \textit{James Webb Space Telescope} (JWST), both in selected photometric bands \citep[e.g.,][]{Zieba2023,Ducrot2024,Xue2024,Bello-Arufe2025} and through photometric phase-curve analyses \citep{Gillon2025}. Earlier observations also include a small number of rocky planets observed in several photometric bands with the \textit{Spitzer} InfraRed Array Camera \citep[IRAC; e.g.,][]{Kreidberg2019}. These datasets represent the first steps toward terrestrial exoplanet characterization, but robust and unambiguous atmospheric detections for rocky planets remain limited. Recent studies have reported atmospheric detections for 55~Cancri~e \citep{Hu2024} and TOI-561b \citep{Teske2025}, but a general observational confirmation across the rocky exoplanet population is still lacking.

A central difficulty is that current observations do not uniquely distinguish secondary atmospheres from airless rocky surfaces. For example, secondary-eclipse measurements near $15\,\mu$m probe strong $\mathrm{CO_2}$ absorption \citep[e.g.,][]{Zieba2023}, while additional photometric coverage at $12.8\,\mu$m provides complementary spectral information that can help distinguish the $15\,\mu$m $\mathrm{CO_2}$ feature from the underlying thermal emission \citep{Ducrot2024}. However, as shown by \citet{Gkouvelis_et_al_2025} and \citet{Hammond2025}, such measurements are not always sufficient to separate atmospheric scenarios from bare-rock interpretations. Phase-curve observations can help reduce this ambiguity \citep{Gillon2025}, since their morphology depends on atmospheric heat redistribution and, in the airless case, on the directional scattering properties of the surface.

The interpretation of these observations relies on forward models spanning both atmospheric and surface scenarios. Atmospheric models are supported by extensive laboratory and theoretical datasets, including molecular opacities and chemical reaction rates over a broad range of temperatures and pressures \citep{Heays2017,Chubb2024,Gkouvelis2024}. As a result, synthetic spectra for atmospheric scenarios can be computed reliably across a wide range of physical regimes \citep[e.g.,][]{Gkouvelis2026ApJ,Gkouvelis2026A&A}. The experimental basis for rocky surface characterization, however, remains substantially more limited.

For airless rocky planets, the dominant observables in reflected light and thermal emission are controlled by the surface reflectance and emissivity. Reflectance and emittance spectroscopy of particulate surfaces builds on decades of work in planetary science \citep{Hapke1981,Lumme1981,Shkuratov2001,Cord2003,Cord2005}. Laboratory bidirectional reflectance distribution function (BRDF) measurements of geological materials are collected in public spectral libraries such as USGS\footnote{\url{https://www.usgs.gov/labs/spectroscopy-lab/usgs-spectral-library}}, RELAB\footnote{\url{https://sites.brown.edu/relab/}}, and others \citep{Pieters2004,Clark2007,Shepard2016,Kokaly2017,Milliken2020,Milliken2021}. These datasets are widely used to interpret remote-sensing observations of planetary surfaces and to construct forward models of rocky exoplanets. Recent laboratory studies of terrestrial magmatic rocks, for example, have used VNIR reflectance spectroscopy to build reference spectral databases for interpreting planetary surfaces such as Mars \citep[e.g.,][]{Barthez2026}, while similar approaches are also applied to disk-integrated observations of asteroids observed with JWST \citep[e.g.,][]{deKleer2026}.

However, most laboratory reflectance libraries provide spectra measured at only a single illumination and viewing geometry, typically $(i,e,g)=(30^{\circ},0^{\circ},30^{\circ})$, where $i$ is the incidence angle, $e$ the emergence angle, and $g$ the phase angle. Existing laboratory measurements nevertheless encompass a broad range of natural geological materials and planetary analogs, including individual minerals, basaltic and other igneous rocks, and powdered or particulate materials representative of planetary regoliths \citep[e.g.,][]{Clark2007,Milliken2021,First2025,Paragas2025}. Laboratory studies have also investigated how surface spectra vary with composition, grain size and texture, temperature, and environmental conditions \citep[e.g.,][]{Helbert2013,Helbert2013b,DonaldsonHanna2017,Ferrari2020,First2025,Fortin2024}. For example, \citet{First2025} measured 15 natural terrestrial basalt samples spanning a range of mineralogies, compositions, and geological settings, while \citet{Paragas2025} extended these measurements to rocky-planet analog materials spanning ultramafic, mafic, intermediate, felsic, and Fe-oxidized compositions, with samples prepared as solid slabs, coarsely crushed material, and fine powders. Thus, laboratory measurements relevant to rocky-planet surfaces span considerable compositional, physical, and environmental diversity. The principal limitation considered here is instead their angular coverage: most measurements do not sample the full bidirectional scattering behavior of the surface. The measurements of \citet{Paragas2025}, for example, were obtained at an incidence angle of $13^{\circ}$. Multi-angular laboratory measurements do exist \citep[e.g.,][]{Shepard2001,Shepard2007}, including measurements of returned Apollo lunar regolith and terrestrial geological materials used as planetary analogs, but remain restricted to a substantially smaller set of materials and spectral ranges.

This limitation has direct consequences for rocky exoplanet characterization. Disk-integrated reflected-light observations depend on the angular scattering behavior of the planetary surface, particularly at small phase angles where opposition effects can strongly enhance the reflected signal. However, laboratory spectra measured at a single illumination and viewing geometry constrain the reflectance at only one point in angular space and therefore do not uniquely determine the scattering behavior required to predict phase curves, geometric albedos, or planet--star contrast ratios (\citet{Gkouvelis2025RNAAS}). As a result, different photometric solutions consistent with the same laboratory spectrum can produce substantially different observable signatures in reflected light. While thermal-emission-dominated regimes are less sensitive to angular scattering behavior, reflected-light observations with JWST, ELTs, the Habitable Worlds Observatory (HWO), and future direct-imaging missions may therefore inherit a systematic uncertainty associated with incomplete laboratory angular constraints.

The exoplanet community has adopted these laboratory libraries to model bare rocky surfaces in forward models and retrieval studies \citep[e.g.,][]{Hu2012,Gkouvelis2025,Hammond2025,Lyu2024}. Additional refinements are currently being explored, both in terms of further physical processes \citep{Lyu2025} and in the treatment of Hapke photometric parameters \citep{Shiltz2023,Gkouvelis2025}. Recent work has also focused on improving the physical interpretation of photometric parameters through microphysical modeling of regolith scattering \citep[e.g.,][]{Xu2026}. However, such approaches typically rely on datasets with sufficient angular coverage. The central limitation is therefore not the sophistication of the photometric model itself, but the incomplete angular coverage of the laboratory datasets used to constrain it.

Large observing programs with JWST are expected to expand the sample of characterized rocky exoplanets, while future facilities such as HWO, the Extremely Large Telescopes (ELTs), and the Large Interferometer For Exoplanets (LIFE) mission concept aim to characterize terrestrial planets in reflected light and thermal emission across a broad range of phase angles. For these facilities, predictive modeling of rocky exoplanet surfaces requires not only spectral information, but also physically realistic angular scattering functions capable of reproducing disk-integrated observables.

In \citet{Helfenstein2025}, strong coupling between Hapke parameters is demonstrated together with the challenges of retrieving unique solutions from disk-integrated phase curves. Their study emphasizes the roles of incomplete phase-angle coverage and observational uncertainties in driving this non-uniqueness. Here, we complement that work by showing that the uncertainty introduced by single-geometry laboratory BRDF measurements is intrinsic to the structure of the forward problem itself. Using the Hapke photometric framework, we demonstrate that multiple combinations of photometric parameters can reproduce the same laboratory reflectance spectrum while predicting substantially different angular scattering behaviors. We further show that this uncertainty propagates directly into disk-integrated exoplanet observables, including reflected-light phase curves, geometric albedos, and planet--star contrast ratios relevant for current and future high-contrast observations. Finally, using multi-angular laboratory measurements from the Bloomsburg University Goniometer (BUG) dataset, we show that additional angular information strongly reduces the allowed parameter space and collapses the uncertainty in predicted observables.

The structure of this paper is as follows. In Section~\ref{sec:photometric_degeneracy}, we describe the origin of the photometric uncertainty associated with single-geometry measurements and its manifestation within the Hapke framework. In Section~\ref{sec:multiangular}, we introduce multi-angular BRDF measurements obtained with the Bloomsburg University Goniometer (BUG)\footnote{\url{https://pds-geosciences.wustl.edu/missions/labdata/}} system \citep{Shepard2001,Shepard2007}. In Section~4, we show how multi-angular measurements constrain the photometric parameter space. In Section~5, we propagate these effects to disk-integrated observables relevant for exoplanet observations, including reflected-light phase curves and geometric albedos. In Section~6, we investigate whether grain-scale optical measurements can mitigate these uncertainties. Finally, in Section~7, we discuss the implications of our results for future rocky exoplanet characterization missions.

\section{Systematic uncertainty from incomplete angular reflectance constraints}
\label{sec:photometric_degeneracy}

\subsection{Hapke model and photometric parameters}

The reflectance of particulate planetary surfaces is commonly described using semi-empirical photometric models such as the Hapke model \citep{Hapke1963,Hapke1981,Hapke1984,Hapke1993,Hapke2021}. In this framework, the bidirectional reflectance distribution function (BRDF) depends on the illumination geometry, the viewing geometry, and a set of parameters describing the scattering properties of the regolith. In the formulation adopted here, the Hapke reflectance is written as

\begin{equation}
r(i,e,g)=
\frac{\omega}{4\pi}
\frac{\mu_0}{\mu_0+\mu}
\left[
(1+B(g))P(g)+H(\mu_0)H(\mu)-1
\right],
\label{eq:hapke}
\end{equation}

where $i$ is the incidence angle, $e$ the emergence angle, and $g$ the phase angle, with $\mu_0=\cos i$ and $\mu=\cos e$.
We describe the single-particle scattering using a single-term Henyey--Greenstein phase function,

\begin{equation}
P(g)=
\frac{1-b^2}
{\left(1+2b\cos g+b^2\right)^{3/2}},
\end{equation}

where $b$ controls the asymmetry of the particle phase function. The shadow-hiding opposition effect (SHOE) is represented by

\begin{equation}
B(g)=
\frac{B_0}
{1+\frac{1}{h}\tan\left(\frac{g}{2}\right)},
\end{equation}

where $B_0$ and $h$ determine the amplitude and angular width of the opposition surge, respectively. Multiple scattering is represented by the Chandrasekhar $H$-function \citep{Chandrasekhar1960,Hapke2012}, for which we use the approximation

\begin{equation}
H(\mu)=
\frac{1+2\mu}
{1+2\mu\sqrt{1-\omega}}.
\end{equation}

The photometric parameters varied in our analysis are therefore the single-scattering albedo $\omega(\lambda)$, the phase-function asymmetry parameter $b$, and the shadow-hiding opposition parameters $B_0$ and $h$. Together, these parameters determine the angular scattering behavior of the surface and therefore control disk-integrated observables such as reflected-light phase curves and geometric albedos.

Additional photometric effects arise from macroscopic surface roughness, which modifies the bidirectional reflectance through large-scale surface slopes \citep{Shepard1998}. Several treatments have been developed to account for these effects \citep{vanDiggelen1959,Veverka1972,Lumme1981,Lumme1981b,Hapke1984,Shiltz2023}. Because the laboratory samples considered in this work are measured on flat surfaces, we neglect macroscopic roughness in the derivation of the photometric parameters from the BRDF measurements. Such effects should, however, be included when modeling disk-integrated observations of real exoplanet surfaces \citep{Gkouvelis2025}.

\begin{figure}
\centering
\includegraphics[width=\columnwidth]{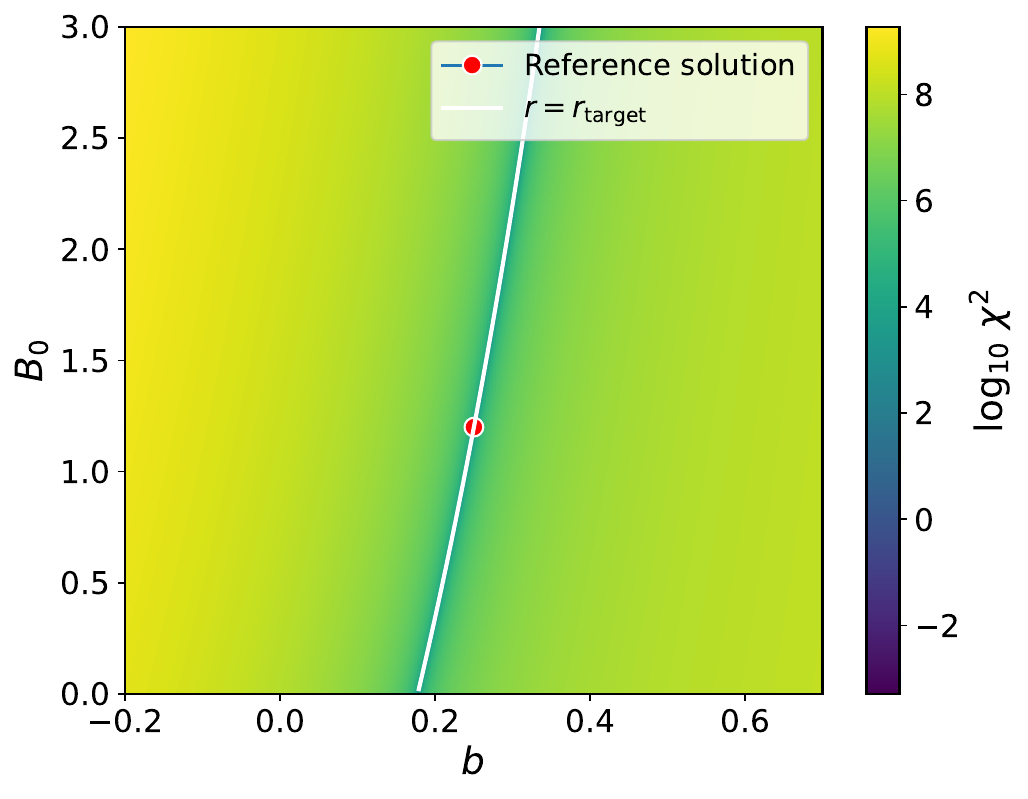}
\caption{
Allowed photometric parameter space in the $(b,B_0)$ plane at fixed laboratory geometry $(i_0,e_0,g_0)=(30^\circ,0^\circ,30^\circ)$. Colors represent the misfit $\log_{10}\chi^2$ between the model reflectance and a reference laboratory reflectance $r_{\rm target}$. Instead of a localized solution, the allowed parameter space forms an extended valley, demonstrating that multiple photometric solutions reproduce the same single-geometry laboratory spectrum. The red point marks the reference parameter combination used to define $r_{\rm target}$; it is not a uniquely recovered solution. The white curve marks the analytic locus defined by $r(i_0,e_0,g_0)=r_{\rm target}$, along which multiple photometric solutions reproduce the same single-geometry reflectance.
}
\label{fig:degeneracy_map}
\end{figure}

\subsection{Single-geometry laboratory measurements}

Laboratory spectral libraries of planetary analog materials typically report bidirectional reflectance spectra measured at a single illumination and viewing geometry, commonly $(i_0,e_0,g_0)=(30^\circ,0^\circ,30^\circ)$ \citep{Clark2007,Milliken2020,Milliken2021}. In this configuration, the measured reflectance at wavelength $\lambda$ can be written as

\begin{equation}
R_{\rm lab}(\lambda)=
r(i_0,e_0,g_0;\omega(\lambda),\boldsymbol{\Theta}),
\end{equation}

where $r$ is the BRDF model and $\boldsymbol{\Theta}$ denotes the set of angular scattering parameters (e.g., $\boldsymbol{\Theta}=\{b,B_0,h\}$). At each wavelength, the laboratory measurement therefore provides only a single constraint on multiple photometric parameters.

The origin of this uncertainty can be understood directly from the structure of the Hapke reflectance. At fixed geometry $(i_0,e_0,g_0)$, the reflected signal depends primarily on the composite quantity

\begin{equation}
K \equiv (1+B(g_0))P(g_0),
\end{equation}

which combines the phase-function and opposition-surge contributions (\citet{Shepard2011}). Different combinations of the photometric parameters can therefore produce the same reflectance at the laboratory geometry while predicting different angular scattering behavior away from that point. As a result, laboratory spectra measured at a single illumination and viewing geometry do not uniquely determine the angular scattering properties required to predict disk-integrated exoplanet observables. Our aim here is deliberately to quantify the range of angular scattering behavior permitted by the laboratory measurement itself, without imposing additional prior assumptions about the surface phase function. Empirical phase functions measured for Solar System surfaces could, in principle, provide physically motivated priors and would reduce the range of allowed solutions, while simpler prescriptions such as Lambertian scattering provide useful reference cases. However,  this approach would introduce angular information that is not contained in the single-geometry laboratory spectrum. The purpose of the present analysis is therefore not to demonstrate the known degeneracy of the Hapke parameters, but to quantify how incomplete angular information in commonly available laboratory measurements propagates into uncertainties in predicted rocky-exoplanet observables.

This limitation has direct observational consequences. Reflected-light observations of rocky exoplanets depend on the angular redistribution of scattered radiation across the planetary disk, particularly at small phase angles where opposition effects can strongly enhance the reflected signal. Different photometric solutions consistent with the same laboratory spectrum can therefore produce substantially different reflected-light phase curves, geometric albedos, and planet--star contrast ratios.

To illustrate this effect, we consider a representative wavelength of $\lambda=0.55\,\mu$m, corresponding to the center of the V band where reflected light dominates over thermal emission for rocky surfaces. We adopt a representative reflectance value $r^\star$ corresponding to a basaltic regolith analog measured at the reference geometry $(i_0,e_0,g_0)=(30^\circ,0^\circ,30^\circ)$.

\begin{figure}
\centering
\includegraphics[width=\columnwidth]{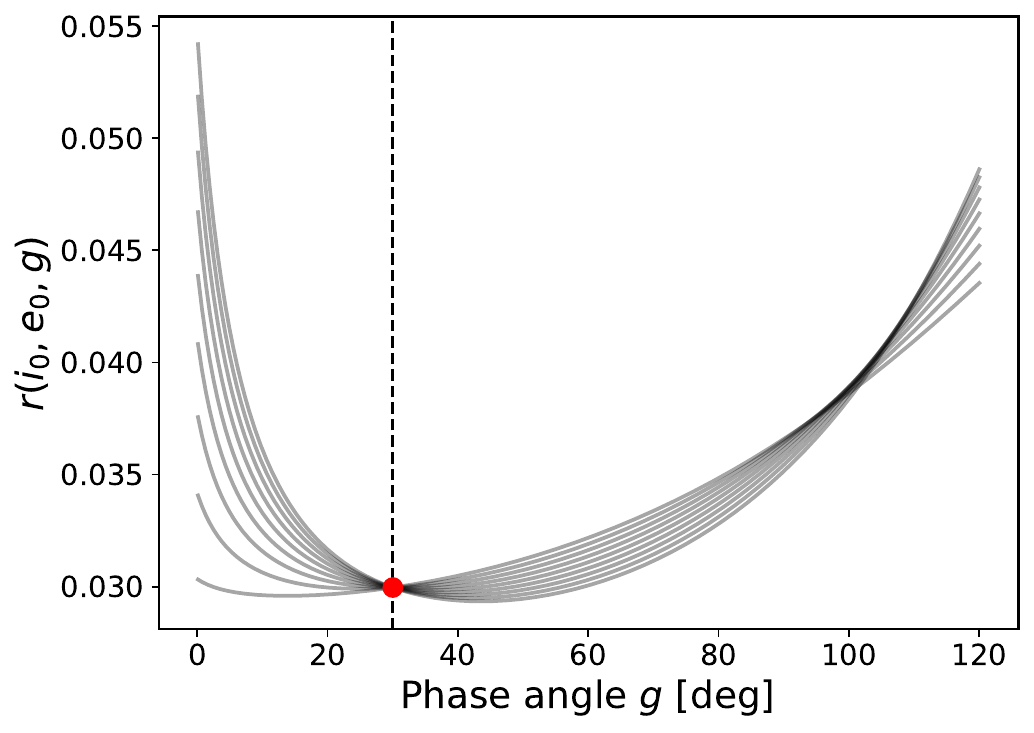}
\caption{
Phase curves corresponding to photometric parameter combinations $(b,B_0)$ that reproduce the same laboratory reflectance at the reference geometry $(i_0,e_0,g_0)=(30^\circ,0^\circ,30^\circ)$. The parameters are explored over $-0.3\leq b\leq0.7$ and $0\leq B_0\leq3$, while the single-scattering albedo and opposition-surge width are fixed at $\omega=0.6$ and $h=0.1$, respectively. All solutions intersect at the measured phase angle $g_0=30^\circ$ by construction, but diverge away from the laboratory geometry, particularly toward small phase angles where opposition effects dominate. This demonstrates that identical laboratory reflectance measurements can correspond to substantially different angular scattering behavior.}

\label{fig:phase_divergence}
\end{figure}

We explore the parameter space $(b,B_0)$ over ranges motivated by previous studies of particulate surfaces \citep[e.g.,][]{Hapke2012}, adopting $-0.3\leq b\leq0.7$ and $0\leq B_0\leq3$. The single-scattering albedo is fixed to $\omega=0.6$, representative of moderately reflective basaltic materials in the visible, while the opposition width is fixed to $h=0.1$.

The misfit is defined as

\begin{equation}
\chi^2(b,B_0)=
\left(
\frac{
r(i_0,e_0,g_0;b,B_0)-r^\star
}{\sigma}
\right)^2,
\end{equation}

where $\sigma$ represents a representative uncertainty. Because the reflectance is evaluated at a single geometry, this $\chi^2$ function is used here primarily as a diagnostic tool to visualize the structure of the allowed parameter space rather than as a formal statistical likelihood.

Figure~\ref{fig:degeneracy_map} shows the resulting misfit surface in the $(b,B_0)$ plane. Instead of a localized minimum, the solution occupies an extended valley, demonstrating that multiple combinations of phase-function asymmetry and opposition-surge amplitude reproduce the same laboratory reflectance. This behavior reflects the fact that the laboratory measurement constrains only the composite scattering contribution at the reference geometry rather than the full angular scattering function.

To quantify the observational implications of this uncertainty, we propagate representative photometric solutions along the allowed parameter manifold into the corresponding reflected-light observables. One corresponds to a strongly backscattering surface with a large opposition-surge amplitude, characteristic of fine-grained porous regoliths exhibiting pronounced coherent backscattering. Another corresponds to a more isotropic or weakly backscattering surface with a smaller opposition contribution. Although these solutions reproduce the same laboratory reflectance at $g_0=30^\circ$, they predict substantially different scattering behavior at other phase angles.

This behavior is illustrated in Fig.~\ref{fig:phase_divergence}. All solutions intersect at the laboratory measurement geometry by construction, but diverge strongly away from that point, particularly toward small phase angles where opposition effects dominate the reflected signal. Thus, identical laboratory spectra can correspond to different reflected-light observables once propagated to disk-integrated exoplanet signals.

More generally, single-geometry laboratory spectra constrain only a projection of the BRDF at one point in angular space rather than the full angular scattering function. The resulting uncertainty is therefore not primarily a consequence of measurement noise or limited precision, but an intrinsic limitation associated with incomplete angular coverage of the laboratory datasets.

\begin{figure}
\centering
\includegraphics[width=\columnwidth,, trim=4cm 0cm 5cm 0cm, clip]{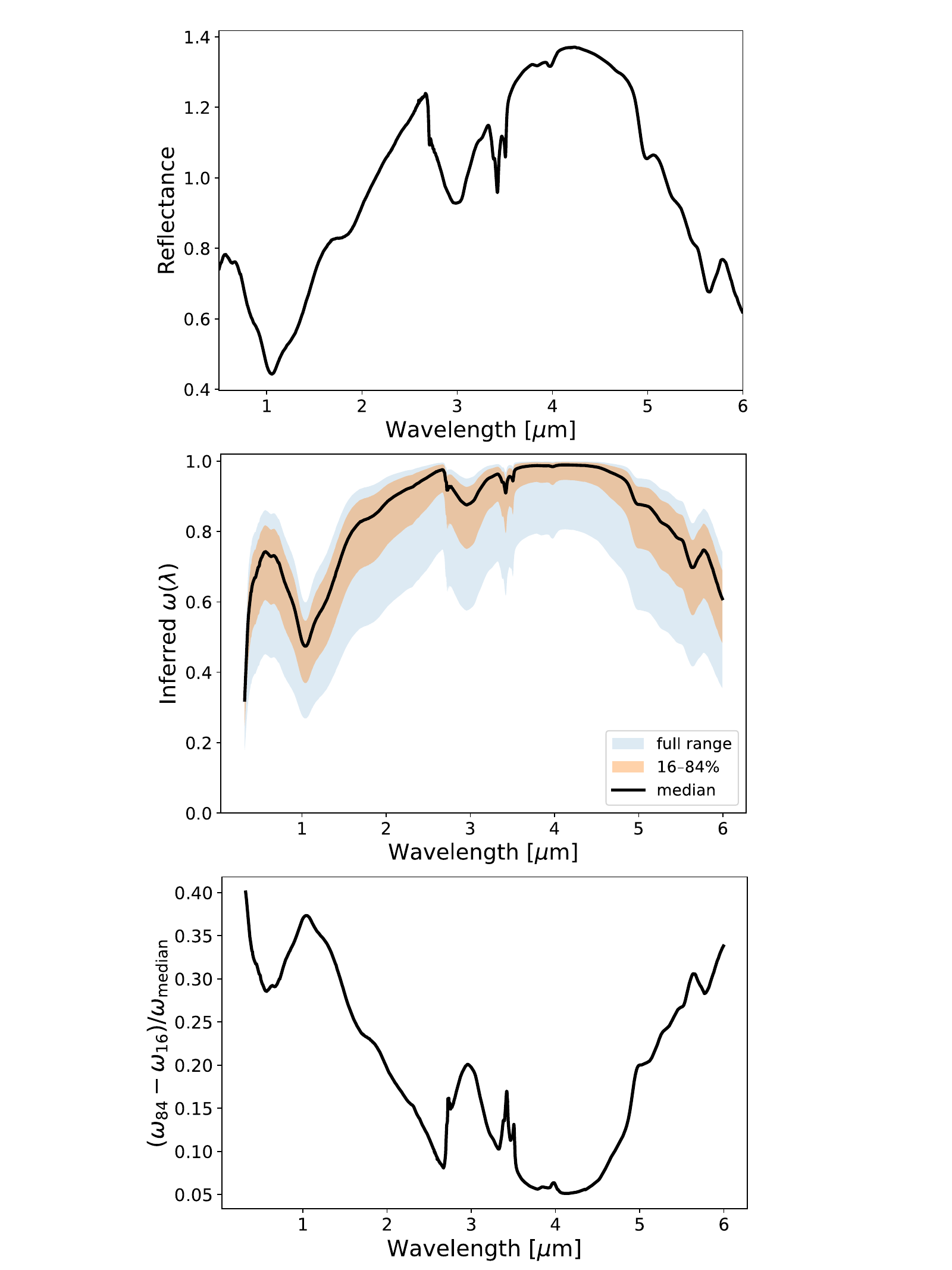}
\caption{
Propagation of incomplete angular constraints into the inferred single-scattering albedo spectrum for the RELAB spectrum BIR1DD065 of olivine from the Martian meteorite NWA~2737. Top: laboratory reflectance spectrum used as the input constraint. Middle: ensemble of inferred $\omega(\lambda)$ solutions obtained by sampling the angular-scattering parameters $(b,B_0,h)$ and inverting the Hapke model wavelength-by-wavelength to reproduce the measured reflectance at the laboratory geometry $(i_0,e_0,g_0)=(30^\circ,0^\circ,30^\circ)$. The solid line shows the median, the darker shaded region encloses the 16th--84th percentile interval, and the lighter shaded region shows the full range of solutions. Bottom: fractional 16th--84th percentile spread in the inferred single-scattering albedo, $(\omega_{84}-\omega_{16})/\omega_{\rm median}$.
}
\label{fig:omega_single_geometry}
\end{figure}

Finally, the uncertainty associated with incomplete angular constraints also propagates into the inferred single-scattering albedo spectrum itself. This effect is illustrated in Fig.~\ref{fig:omega_single_geometry} using the RELAB spectrum BIR1DD065 of olivine from the Martian meteorite NWA~2737 over the wavelength range $0.5$--$6\,\mu$m. The laboratory reflectance spectrum used as the input constraint is shown in the top panel. To quantify the range of single-scattering albedo spectra compatible with this single-geometry measurement, we uniformly sample the angular-scattering parameters over $-0.6\leq b\leq0.6$, $0\leq B_0\leq5$, and $0.01\leq h\leq0.40$. For each sampled combination $(b,B_0,h)$, we numerically invert the Hapke reflectance equation independently at each wavelength to determine the value of $\omega(\lambda)$ that reproduces the measured reflectance at the laboratory geometry $(i_0,e_0,g_0)=(30^\circ,0^\circ,30^\circ)$. The resulting ensemble of $\omega(\lambda)$ solutions is shown in the middle panel, with the solid line denoting the median, the darker shaded region the 16th--84th percentile interval, and the lighter region the full range of solutions. The bottom panel quantifies the corresponding fractional spread, $(\omega_{84}-\omega_{16})/\omega_{\rm median}$. This procedure is not an optimization for a unique best-fitting photometric solution; rather, it maps the family of $\omega(\lambda)$ solutions compatible with the same single-geometry laboratory measurement when the angular-scattering parameters are not independently constrained. The resulting spread therefore directly quantifies the uncertainty in the inferred optical properties introduced by incomplete angular coverage.

Those uncertainties have direct effects on the physical interpretation of the inferred surface properties. The single-scattering albedo $\omega(\lambda)$ describes the balance between scattering and absorption within the regolith, and its spectral dependence carries information about the optical properties of the surface material. Uncertainty in $\omega(\lambda)$ can therefore modify the inferred strength and contrast of spectral features and comparisons with candidate surface compositions. Thus, even when a single-geometry laboratory reflectance spectrum is reproduced exactly, the inferred optical properties of the corresponding surface are not uniquely determined unless its angular scattering behavior is independently constrained.

\section{Multi-angular laboratory constraints}
\label{sec:multiangular}

\subsection{BUG dataset}

To investigate whether additional angular information can reduce the uncertainties described in Section~2, we use laboratory multi-angular reflectance measurements from the Bloomsburg University Goniometer (BUG) facility \citep{Shepard2001,Shepard2007,Shepard2011}. The BUG instrument uses a computer-controlled goniometer to measure bidirectional reflectance over a wide range of illumination and viewing geometries. Incidence angles extend to $60^\circ$ and emergence angles to $80^\circ$, providing measurements over phase angles of approximately $3^\circ$--$140^\circ$. The measurements are calibrated against a Spectralon reference standard and reported as radiance factors.

Here we consider BUG measurements of returned Apollo lunar soil \citep{ShepardPDS2016} and Mars soil analogs \citep{ShepardWardPDS2006}. The Apollo archive contains six returned lunar soil samples from the Apollo~11, 12, 15, 16, and 17 missions, rather than lunar simulants. In particular, the retrieval analysis presented below uses Apollo~12 soil sample 12001. The Apollo samples were measured at four wavelength bands centered at 453, 550, 751, and 952~nm. The Mars archive contains five soil analogs (AREF~235, AREF~236, AREF~237, AREF~238, and JSC~Mars-1), measured using flight-spare filters corresponding to the Mars Exploration Rover Pancam wavelength range. For each measurement, the archive provides the incidence, emergence, azimuth, and phase angles together with the corrected radiance factor. Unlike most laboratory spectral libraries, which provide reflectance measurements at only a single illumination--viewing geometry, the BUG measurements densely sample the bidirectional scattering behavior across hundreds of angular configurations. The data are publicly available through the NASA Planetary Data System (PDS) Geosciences Node\footnote{\url{https://pds-geosciences.wustl.edu/missions/labdata/}}.

\begin{figure}
\centering
\includegraphics[scale=0.42, trim=6.6cm 0.1cm 3cm 0cm, clip]{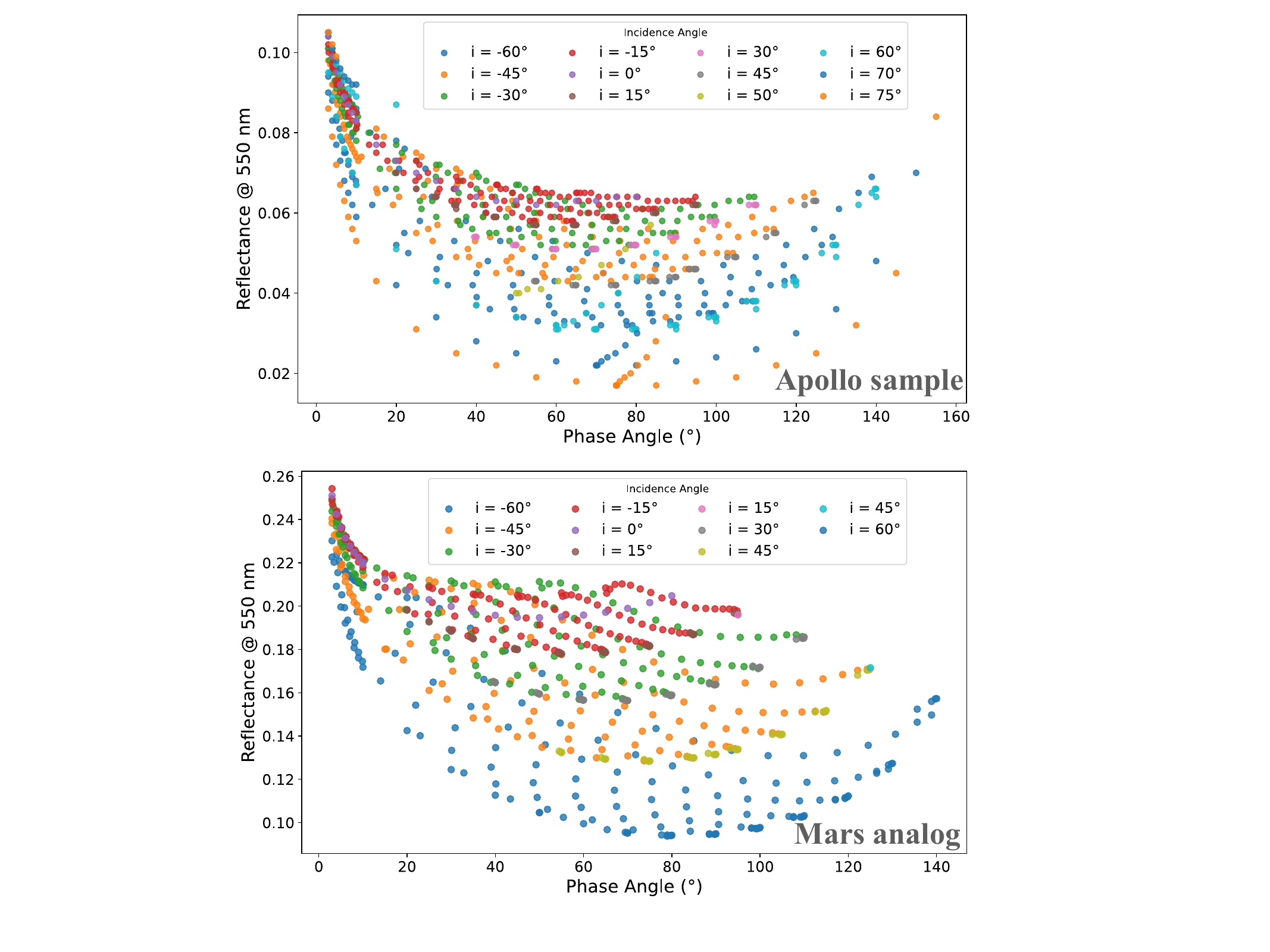}
\caption{
Representative multi-angular bidirectional reflectance measurements from the BUG dataset for a returned Apollo lunar soil and a Mars soil analog. The broad angular coverage samples a wide range of incidence and emergence geometries, directly constraining the phase-angle dependence of particulate surface scattering. In contrast to single-geometry laboratory spectra, such datasets provide direct information on the angular scattering behavior required to model disk-integrated reflected-light observations of rocky exoplanets.
}
\label{fig:rock}
\end{figure}

Figure~\ref{fig:rock} shows representative multi-angular measurements for a returned Apollo lunar soil and a Mars soil analog. The dense angular coverage reveals systematic variations in reflectance with phase angle that are inaccessible to single-geometry laboratory spectra. In particular, the strong increase in brightness toward small phase angles directly constrains the opposition behavior of particulate surfaces, which strongly influences reflected-light exoplanet observables near opposition.

\subsection{Bayesian retrieval framework}

To quantify the impact of angular diversity on the inferred surface photometry, we perform Bayesian inversions of the Hapke model using the BUG reflectance measurements.

For each measurement geometry $(i,e,g)$, the forward model is the Hapke bidirectional-reflectance function introduced in Eq.~\ref{eq:hapke},
\begin{equation}
r_i^{\rm model}
=
r(i_i,e_i,g_i;\boldsymbol{\theta}),
\end{equation}
where
\begin{equation}
\boldsymbol{\theta}=(\omega,b,B_0,h)
\end{equation}
is the vector of retrieved photometric parameters. Here, $\omega$ is the single-scattering albedo, $b$ controls the particle phase function, and $(B_0,h)$ parameterize the amplitude and angular width of the shadow-hiding opposition effect, respectively.

The posterior probability distribution of the photometric parameters is sampled according to

\begin{equation}
P(\omega,b,B_0,h \mid D)
\propto
\mathcal{L}(D \mid \omega,b,B_0,h)\,
P(\omega,b,B_0,h),
\end{equation}

where $D$ denotes the measured reflectances across all sampled geometries and $P(\omega,b,B_0,h)$ represents the prior distribution of the parameters.

We adopt uniform priors over the ranges

\[
0.1 \leq \omega \leq 0.9,
\]
\[
-0.5 \leq b \leq 0.8,
\]
\[
0 \leq B_0 \leq 5,
\]
\[
0.01 \leq h \leq 0.8.
\]

These intervals are motivated by previous studies of planetary regolith scattering \citep[e.g.,][]{Hapke2012}. In particular, the explored range of the phase-function asymmetry parameter $b$ encompasses both backscattering and forward-scattering regimes typically inferred for Solar System surfaces, while the allowed range of $B_0$ spans the opposition-surge amplitudes commonly observed in laboratory and planetary measurements. These ranges are deliberately broad and are intended to encompass the scattering behavior of the particulate planetary surfaces considered here rather than to impose restrictive prior knowledge on a particular material. As a direct example, the multi-angular retrieval of Apollo~12 soil sample 12001 at 550~nm yields $\omega\simeq0.58$, $b\simeq-0.08$, $B_0\simeq4.15$, and $h\simeq0.08$ (Table~\ref{tab:retrieval_parameters}), all within the adopted parameter space. The broad prior ranges therefore encompass the solution inferred directly from the angular scattering behavior of the returned lunar soil while allowing substantially different scattering behaviors to be explored when only a single laboratory geometry is available.

Assuming Gaussian measurement uncertainties, the likelihood function is

\begin{equation}
\ln \mathcal{L}
=
-\frac{1}{2}
\sum_i
\left(
\frac{
r_i^{\rm obs}-r_i^{\rm model}
}{
\sigma_i
}
\right)^2,
\end{equation}

where $r_i^{\rm obs}$ and $r_i^{\rm model}$ denote the observed and model-predicted reflectances at geometry $i$, and $\sigma_i$ is the corresponding measurement uncertainty. We adopt a representative relative uncertainty of $\sim5\%$, consistent with typical laboratory BRDF measurements.

To isolate the role of angular diversity, we perform two separate inversions:

\begin{enumerate}
\item a single-geometry inversion using only measurements at the reference geometry $(i_0,e_0,g_0)$,
\item a multi-angular inversion using the full BUG dataset.
\end{enumerate}

The posterior distributions are sampled using the \texttt{emcee} Markov Chain Monte Carlo ensemble sampler \citep{Foreman2013}. We use 100 walkers and run the chains for $10^4$ steps, discarding an initial burn-in phase and verifying convergence through inspection of the chain evolution and autocorrelation times.

The goal of these retrievals is not primarily to obtain a unique ``best-fit'' photometric solution, but to quantify how incomplete angular coverage propagates into uncertainty in the inferred scattering properties of the surface. By comparing the single-geometry and multi-angular inversions, we directly evaluate how angular diversity constrains the photometric parameter space and reduces uncertainty in reflected-light observables relevant for rocky exoplanet characterization.

\section{Reduction of photometric uncertainty with multi-angular data}
\label{sec:degeneracy_breaking}

\begin{figure*}
\centering
\includegraphics[scale=0.52, trim=0cm 5.5cm 0cm 3cm, clip]{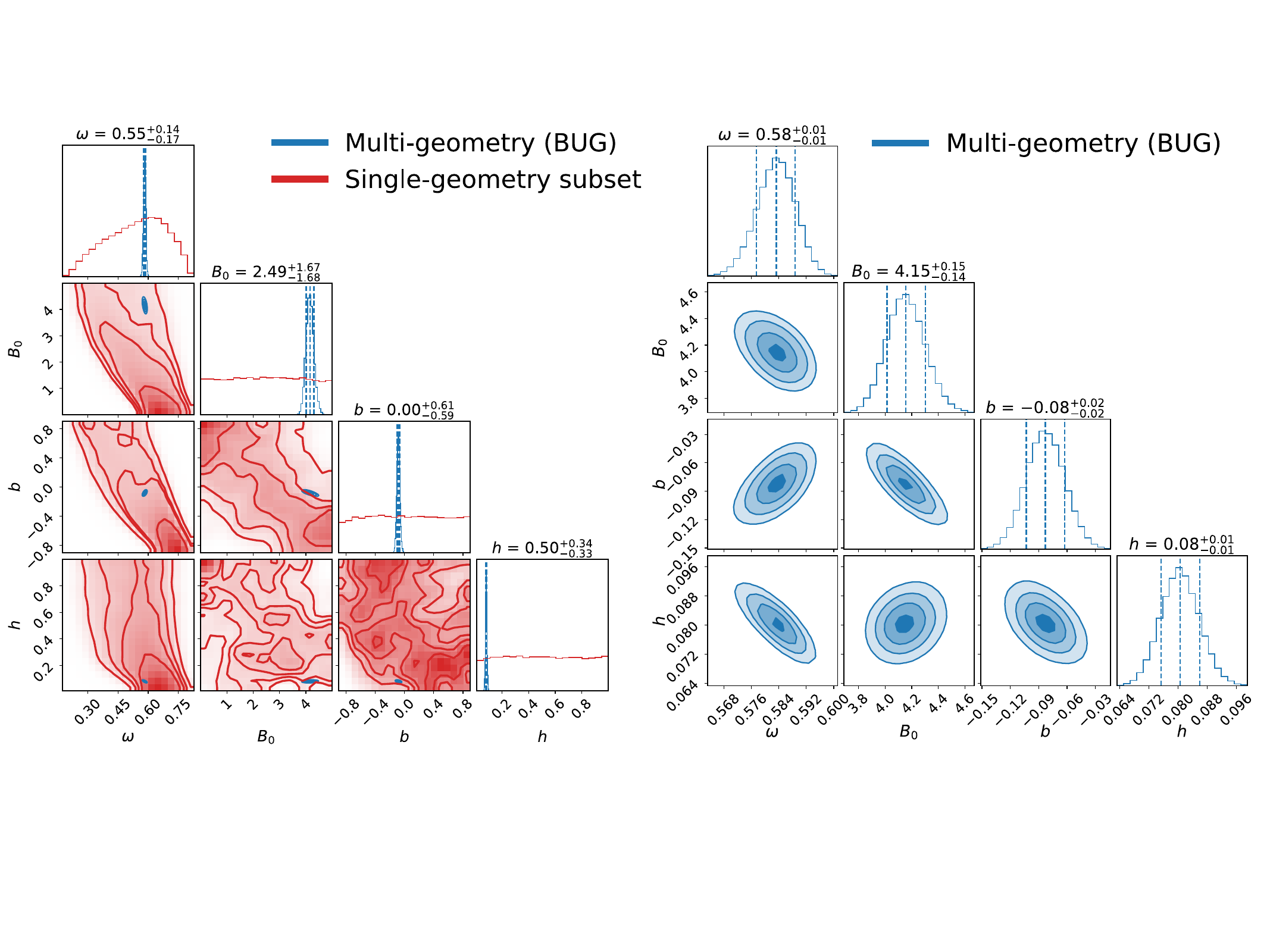}
\caption{Posterior distributions of the Hapke parameters for Apollo~12 soil sample 12001 at 550~nm. The left panel compares the retrieval using the restricted single-geometry subset (red) with that using the full multi-geometry BUG dataset (blue). The values above the diagonal panels indicate the median and 16th--84th percentile credible intervals for the single-geometry retrieval. The right panel shows the multi-geometry posterior on an expanded scale, with the corresponding median and 16th--84th percentile credible intervals. The much broader single-geometry posterior illustrates the strong parameter degeneracies introduced by limited angular sampling.}
\label{fig:corner}
\end{figure*}

Figure~\ref{fig:corner} compares the posterior distributions obtained from the single-geometry and multi-angular inversions of the Hapke model parameters. When only a single illumination-viewing geometry is used, the posterior occupies an extended region of parameter space, reflecting strong correlations between the single-scattering albedo and the angular scattering parameters. In contrast, conditioning the inversion on the full multi-angular BUG dataset collapses the posterior to a compact region, substantially reducing the uncertainty in the inferred scattering properties of the surface.

For the representative Apollo~12 soil sample 12001 at 550~nm considered here, Table~\ref{tab:retrieval_parameters} summarizes the posterior medians and 68\% credible intervals obtained from the single-geometry and multi-angular retrievals.

\begin{table}
\centering
\caption{Posterior constraints on the Hapke parameters for Apollo~12 soil sample 12001 at 550~nm. Values correspond to the posterior median and 16th--84th percentile credible interval. The uncertainty-reduction factor is defined as the ratio of the 16th--84th percentile widths of the single-geometry and multi-angular posteriors.}
\label{tab:retrieval_parameters}
\begin{tabular}{lccc}
\hline
Parameter & Single & Multi-angular & Reduction factor \\
\hline
$\omega$ & $0.55^{+0.14}_{-0.17}$ & $0.583^{+0.006}_{-0.006}$ & 27.6 \\
$B_0$    & $2.49^{+1.67}_{-1.68}$ & $4.15^{+0.15}_{-0.14}$ & 11.6 \\
$b$      & $0.00^{+0.61}_{-0.59}$ & $-0.083^{+0.021}_{-0.021}$ & 29.3 \\
$h$      & $0.50^{+0.34}_{-0.33}$ & $0.081^{+0.005}_{-0.005}$ & 63.1 \\
\hline
\end{tabular}
\end{table}

Using the width of the 16th--84th percentile interval as a measure of uncertainty, the inclusion of multi-angular constraints reduces the credible intervals by factors of 11.6--63.1 for the four photometric parameters considered here (Table~\ref{tab:retrieval_parameters}). The strongest reduction occurs for the opposition-surge width $h$, whose credible interval narrows by a factor of 63.1, while the uncertainties in $\omega$, $B_0$, and $b$ decrease by factors of 27.6, 11.6, and 29.3, respectively. 

The comparison shown in Fig.~\ref{fig:corner}, using Apollo~12 soil sample 12001 at 550~nm, demonstrates that the large uncertainty envelope obtained from single-geometry laboratory measurements is not primarily an intrinsic limitation of the Hapke formalism itself, but rather a consequence of incomplete angular information in the experimental datasets used to constrain the model.

When the reflectance is sampled across a sufficiently broad range of incidence, emergence, and phase angles, the angular dependence of the scattering function becomes directly constrained by the laboratory measurements. The resulting posterior distributions collapse to a localized region of parameter space with well-defined parameter correlations. The remaining posterior volume therefore represents the physically allowed range of photometric solutions consistent with the measured angular scattering behavior of the surface.

Importantly, the reduction in parameter uncertainty is not merely a formal improvement in the inversion. The opposition-surge parameters and phase-function asymmetry directly control the redistribution of reflected light across phase angle and therefore determine the disk-integrated brightness of rocky exoplanets in reflected light. As a consequence, the reduction of photometric uncertainty shown in Fig.~\ref{fig:corner} propagates directly into improved predictions of reflected-light phase curves, geometric albedos, and planet--star contrast ratios.

These results demonstrate that the dominant limitation in current rocky exoplanet surface models is not necessarily the complexity of the photometric formalism, but the incomplete angular coverage of the laboratory datasets used to constrain it. Multi-angular laboratory measurements provide the missing information required to construct physically predictive scattering models for rocky exoplanet observations.

In the following section, we propagate these uncertainties into disk-integrated observables relevant for current and future reflected-light exoplanet observations.

\begin{figure}
\centering
\includegraphics[width=\columnwidth]{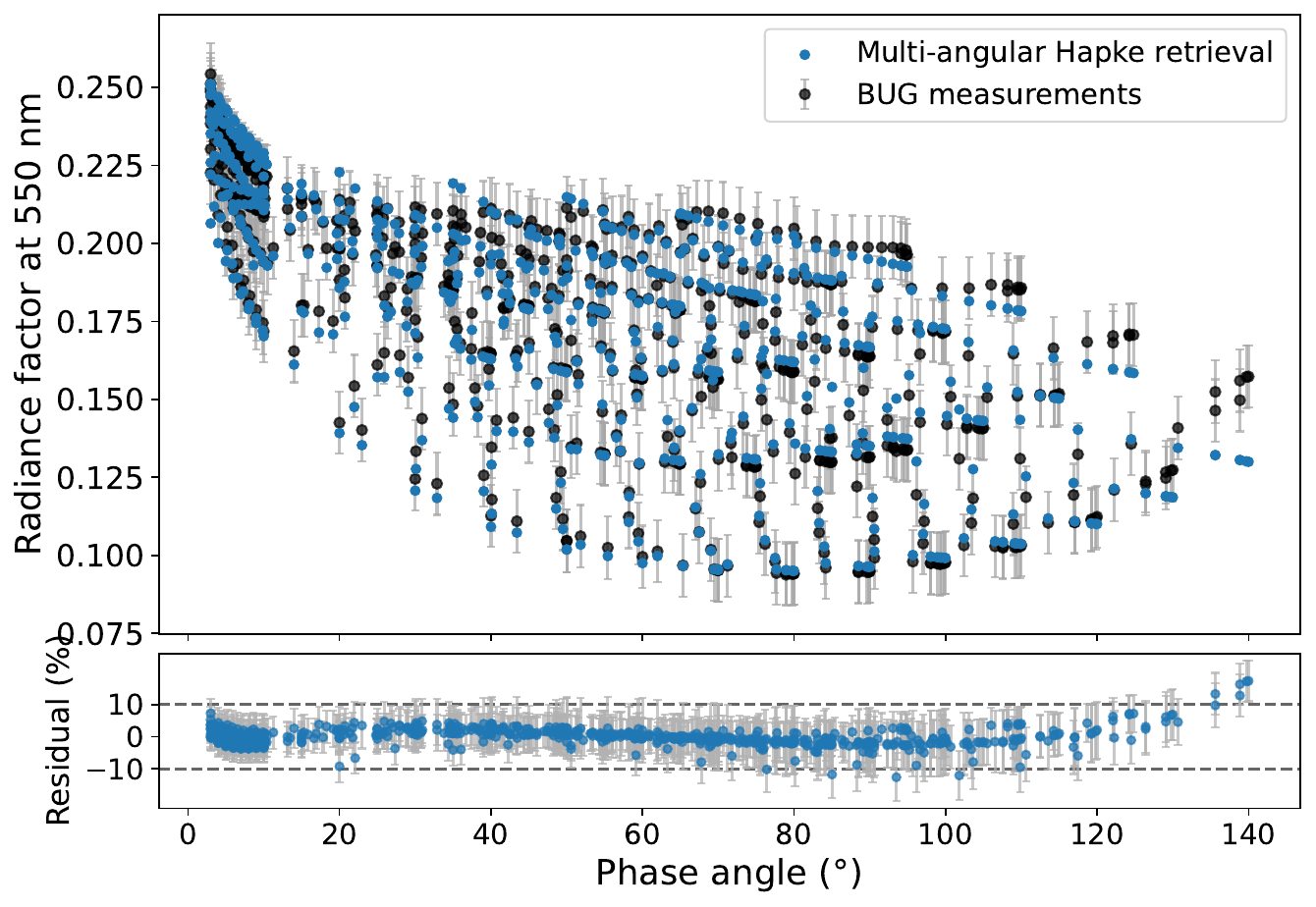}
\caption{
Comparison between the BUG multi-angular measurements and the retrieved Hapke model for the Mars analog at 550~nm. Top: measured radiance factors and predictions from the posterior-median multi-angular retrieval, evaluated at the same incidence, emergence, and phase angles as each laboratory measurement. Bottom: fractional residuals, defined as $100(r_{\rm obs}-r_{\rm model})/r_{\rm obs}$. The dashed horizontal lines indicate $\pm10\%$ residuals. The comparison provides a direct assessment of the ability of the retrieved photometric solution to reproduce the measured angular scattering behavior.
}
\label{fig:retrieval_fit}
\end{figure}

\begin{figure}[h!]
\centering
\includegraphics[scale=0.41, trim=3.4cm 0.5cm 5cm 0cm, clip]{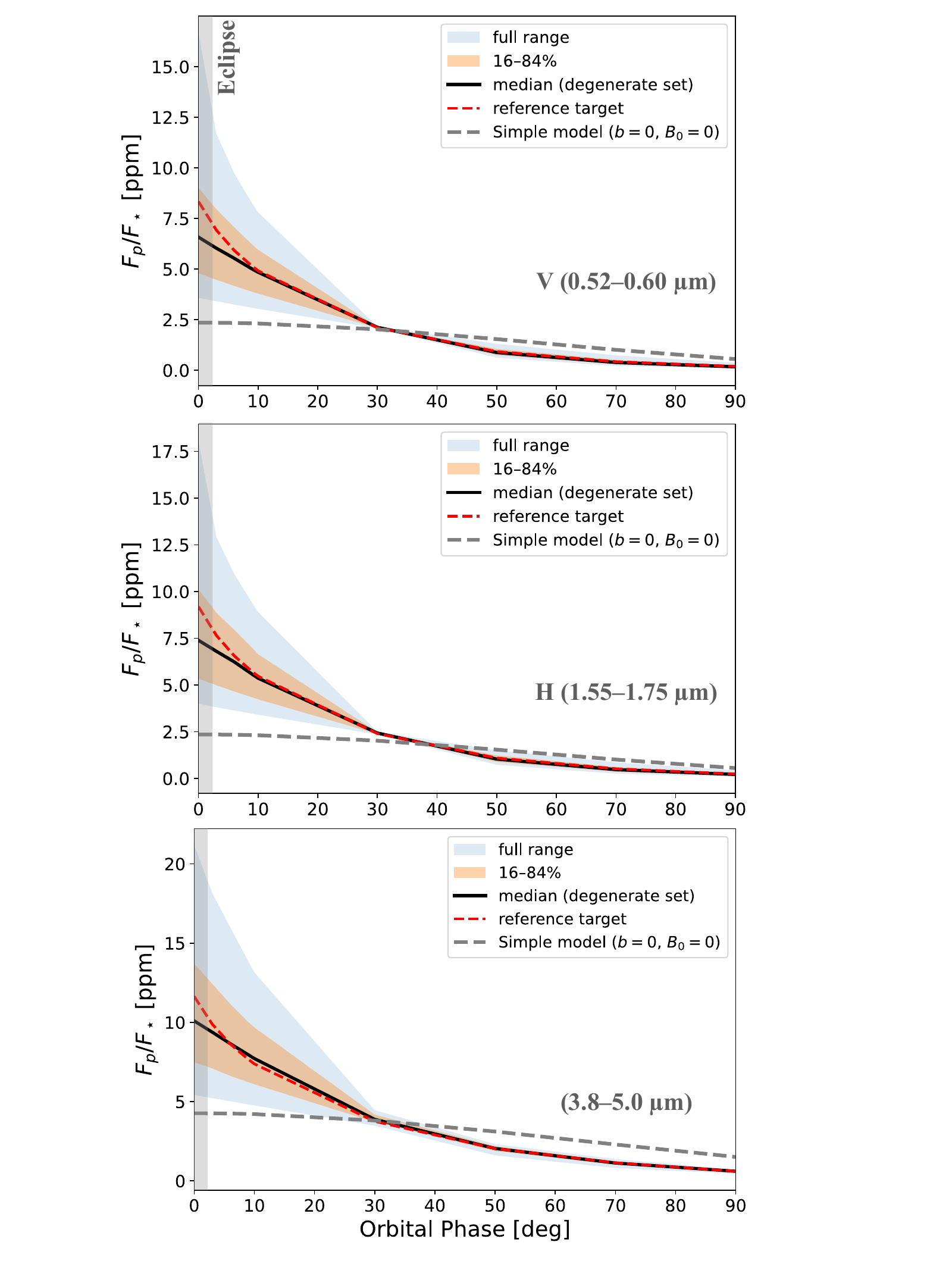}
\caption{
Band-integrated phase curves for three representative wavelength regimes: reflection-dominated (0.52--0.60~$\mu$m), mixed reflection--thermal (1.55--1.75~$\mu$m), and thermal-dominated (3.8--5.0~$\mu$m). The shaded regions indicate the range of disk-integrated phase curves predicted by photometric parameter sets consistent with the same single-geometry laboratory reflectance spectrum. The darker shaded region corresponds to the central 68\% interval, while the lighter region shows the full extent of the solutions. The black solid line represents the median predicted phase curve, and the red dashed line corresponds to the reference laboratory reflectance behavior. The gray dashed line shows the prediction of a simplified photometric model assuming isotropic scattering ($b=0$) and no opposition effects ($B_0=0$), with the single-scattering albedo adjusted to reproduce the laboratory reflectance at the reference geometry. Although this simplified model matches the laboratory measurement by construction, it produces substantial systematic biases in the predicted reflected-light observables away from the measurement geometry.
}
\label{fig:fraction}
\end{figure}

\section{Propagation to disk-integrated exoplanet observables}
\label{sec:phasecurve_propagation}

We now propagate the photometric uncertainty associated with incomplete angular laboratory constraints into disk-integrated observables relevant for rocky exoplanet observations. Disk-integrated phase curves have been measured for a wide range of Solar System bodies, including Saturnian satellites, Pluto, Charon, Triton, dwarf planets, and Kuiper Belt objects, using both spacecraft and ground-based observations \citep[e.g.,][]{Hedman2010,Olkin2021,Howett2021,Verbiscer2022,Helfenstein2025}. Such measurements provide empirical examples of how the angular scattering properties of unresolved planetary surfaces are encoded in their whole-disk phase-dependent brightness.
In such disk-integrated measurements, the observed signal represents an integral over the full distribution of scattering geometries across the illuminated planetary disk. As a consequence, uncertainty in the angular scattering behavior of the surface propagates directly into uncertainty in observable quantities such as reflected-light phase curves, geometric albedos, and planet--star contrast ratios.

To quantify this effect, we compute synthetic reflected-light phase curves using the Hapke parameter distributions obtained from the retrievals described in previous sections. For each sampled parameter set $(\omega,b,B_0,h)$, we evaluate the disk-integrated reflected flux

\begin{equation}
\frac{F_p}{F_\star}(\lambda,\alpha)
=
\left(
\frac{R_p}{a}
\right)^2
\iint_{\rm visible}
r(i,e,\alpha;\omega,b,B_0,h)\,
\mu_0\,\mu\,d\Omega,
\label{eq:disk}
\end{equation}

where $R_p$ is the planetary radius, $a$ the orbital separation, $\alpha$ the phase angle, and $d\Omega$ the surface element on the planet. The integral is performed over the region simultaneously illuminated by the star and visible to the observer \citep[e.g.,][]{Shkuratov2001,Hu2012,Gkouvelis2025}. The quantities $\mu_0=\cos i$ and $\mu=\cos e$ denote the directional cosines of the illumination and viewing angles, respectively.

Each parameter sample therefore produces a predicted disk-integrated phase curve. The ensemble of solutions defines an uncertainty envelope corresponding to the range of reflected-light observables consistent with the laboratory constraints.

Figure~\ref{fig:fraction} shows the resulting phase-curve envelopes for three representative wavelength bands: a reflection-dominated band (0.52--0.60~$\mu$m), a mixed reflection--thermal band (1.55--1.75~$\mu$m), and a thermal-dominated band (3.8--5.0~$\mu$m). The phase-curve envelopes are constructed by sampling several hundred parameter sets from the posterior distribution obtained from the single-geometry inversion. For each sample, the disk-integrated phase curve is computed using Eq.~\ref{eq:disk}, and the resulting ensemble defines the uncertainty envelope shown in Fig.~\ref{fig:fraction}.

In the reflection-dominated band (0.52--0.60~$\mu$m), the predicted phase curves exhibit a large spread at small phase angles. Backscattering-dominated solutions produce enhanced brightness near opposition, while more isotropic scattering solutions yield significantly weaker reflected-light signals. Although these solutions reproduce the same laboratory reflectance at the reference geometry, they diverge strongly once propagated to disk-integrated observables.

At intermediate wavelengths (1.55--1.75~$\mu$m), where both reflected light and thermal emission contribute to the observed signal, the uncertainty envelope becomes moderately reduced because part of the flux is controlled by thermal emission. Thermal emission depends primarily on the hemispherical emissivity of the surface, which is related to the integrated reflectance rather than its detailed angular redistribution, and is therefore less sensitive to uncertainty in the photometric scattering parameters.

In the thermal-dominated band (3.8--5.0~$\mu$m), the sensitivity to angular scattering behavior decreases further, and the spread in the predicted phase curves becomes correspondingly smaller. In this regime, the observable signal is governed primarily by surface emissivity rather than directional scattering.

We quantify the impact of incomplete angular constraints as a function of phase angle across the three wavelength regimes. At small phase angles ($\alpha \lesssim 10^\circ$), the fractional uncertainty associated with the photometric solutions remains consistently high in all bands, with $\delta_{\rm model}\simeq0.5$, corresponding to an uncertainty of order 50\% in the reflected-light signal.

In the direct-imaging regime ($\alpha\sim60^\circ$--$90^\circ$), the behavior depends strongly on wavelength. In the reflection-dominated band, the uncertainty remains comparably large, with $\delta_{\rm model}\sim0.5$--$0.6$, indicating that uncertainty in the angular scattering behavior persists even at relatively large phase angles. In contrast, in the mixed and thermal-dominated bands, the uncertainty decreases to $\delta_{\rm model}\sim0.1$--$0.2$, reflecting the decreasing sensitivity of the observable to directional scattering as thermal emission contributes more significantly to the signal.

We additionally evaluate the performance of a simplified photometric model assuming isotropic scattering ($b=0$) and no opposition effects ($B_0=0$). Although this model reproduces the laboratory reflectance at the reference geometry by construction, it introduces substantial systematic biases at other phase angles.

At small phase angles, the simplified model underestimates the reflected flux by approximately 50\%--65\%, reflecting the absence of opposition effects. At larger phase angles ($\alpha\sim60^\circ$--$90^\circ$), the same model systematically overestimates the reflected signal by more than a factor of two across all wavelength regimes.

The corresponding absolute errors in the predicted planet--star contrast are of order $3$--$5$~ppm at small phase angles and $\sim0.4$--$1$~ppm in the direct-imaging regime. These errors are substantial in the context of rocky-exoplanet characterization because they correspond to systematic uncertainties of tens of percent in the predicted planetary signal rather than to a small perturbation of the phase curve. Such uncertainties can directly affect the interpretation of the measured brightness in terms of geometric albedo and surface reflectivity, and can complicate comparisons between observed reflected-light spectra and candidate surface compositions. In all cases, the systematic bias introduced by simplified photometric assumptions is comparable to or larger than the intrinsic uncertainty associated with incomplete angular constraints themselves.

These results demonstrate that incomplete angular laboratory constraints propagate directly into uncertainty and systematic bias in observable rocky exoplanet phase curves and reflected-light brightness. Importantly, this uncertainty is systematic rather than statistical, reflecting the inability of single-geometry laboratory spectra to uniquely constrain the angular scattering behavior of particulate surfaces. As a consequence, even high-precision laboratory reflectance measurements obtained at a single geometry are insufficient to uniquely predict the reflected-light signal of a rocky exoplanet, potentially limiting the interpretation of observed brightness variations in terms of surface albedo, scattering behavior, and composition.

\begin{figure}[h!]
\centering
\includegraphics[width=\columnwidth]{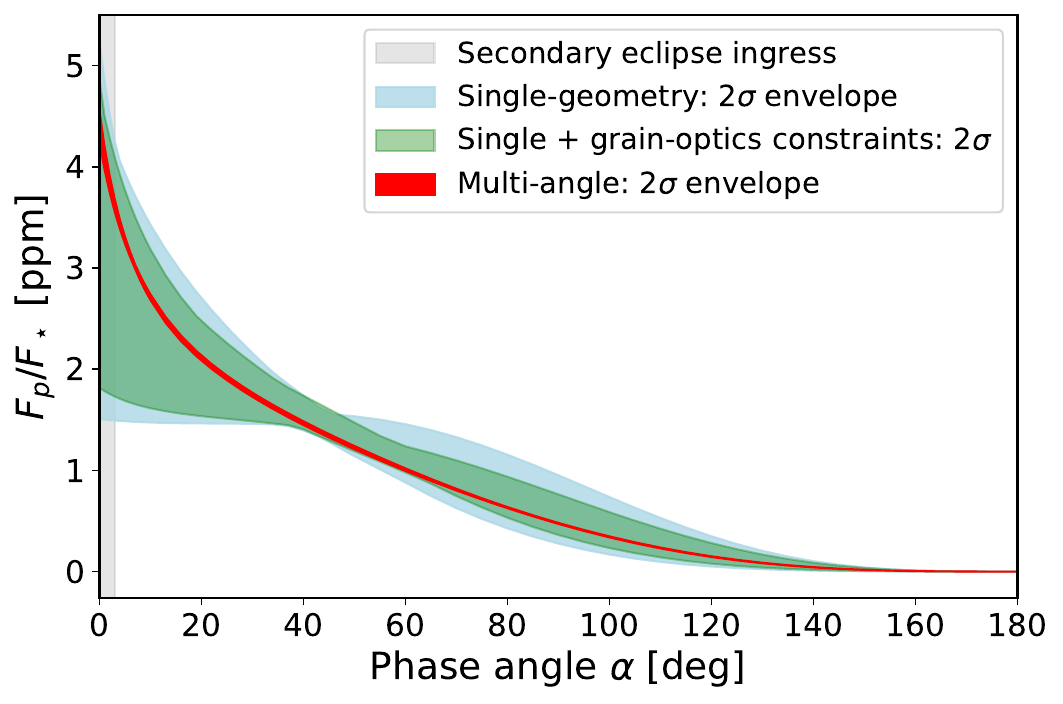}
\caption{
Disk-integrated reflected-light phase curves at $0.55\,\mu$m for a TRAPPIST-1b-like configuration using an Apollo regolith analog (sample 10084) from the BUG dataset \citep{Shepard2007}. The light-blue region shows the uncertainty envelope obtained from single-geometry laboratory conditioning. The green region corresponds to the case where additional grain-scale optical priors constrain $\omega$ and $b$. The red region shows the predictions obtained using the full multi-angular BUG dataset. Shaded regions represent the propagated posterior uncertainty in the Hapke parameters $(\omega,b,B_0,h)$. The vertical gray band marks the approximate phase-angle range corresponding to secondary-eclipse ingress. Although grain-scale optical constraints partially reduce the uncertainty envelope, substantial uncertainty remains until direct multi-angular constraints are introduced.
}
\label{fig:phasecurve_comparison}
\end{figure}

\section{Can indirect constraints replace multi-angular laboratory measurements?}

In the previous sections, we showed that incomplete angular coverage in laboratory reflectance datasets propagates directly into uncertainty in disk-integrated exoplanet observables. A natural question is therefore whether indirect constraints or simplified modeling assumptions can compensate for the absence of direct multi-angular BRDF measurements.

In this section, we investigate three commonly invoked alternatives: (i) constraints from grain-scale optical measurements, (ii) simplified photometric assumptions frequently adopted in exoplanet studies, and (iii) observational diagnostics based on geometric albedo and reflected-light contrast. Our goal is to determine whether these approaches can provide sufficiently predictive reflected-light models in the absence of direct angular laboratory constraints.

\subsection{Constraints from grain-scale optical measurements}
\label{sec:dust_constraints}

Laboratory measurements of planetary analog materials often focus on the optical properties of individual grains or dust aggregates \citep[e.g.,][]{Munoz2010,Munoz2025}. In particular, measurements of complex refractive indices $(n,k)$ and single-particle scattering matrices provide important constraints on the intrinsic optical properties of minerals. Combined with scattering theories such as Mie or T-matrix calculations, these measurements can be used to derive single-particle phase functions and single-scattering albedos.

In principle, such information could help constrain the photometric parameters entering semi-empirical surface reflectance models such as the Hapke formalism. For example, particle-scattering calculations may provide estimates of the single-scattering albedo $\omega$ and the phase-function asymmetry parameter $b$ for a given particle composition and size distribution.

However, grain-scale optical measurements characterize isolated particles rather than the macroscopic scattering behavior of a granular planetary surface. Planetary regoliths consist of densely packed particulate media in which photons undergo multiple scattering, inter-particle shadowing, coherent backscattering, and porosity-dependent radiative transfer. These effects depend not only on the intrinsic optical constants of the material, but also on macroscopic surface properties such as grain packing, porosity, particle-size distribution, and roughness.

To evaluate whether grain-scale constraints can reduce the uncertainty identified in previous sections, we perform a numerical experiment. Starting from the parameter space obtained from the single-geometry inversion (Section~\ref{sec:degeneracy_breaking}), we introduce external priors on $\omega$ and $b$ motivated by Apollo regolith analog measurements from the BUG dataset \citep{Shepard2007}. Specifically, we restrict $\omega$ and $b$ to ranges consistent with plausible grain-scattering behavior while leaving the opposition-surge parameters $(B_0,h)$ unconstrained.

Using these constrained parameter sets, we compute disk-integrated phase curves following the procedure described in Section~\ref{sec:phasecurve_propagation}. Figure~\ref{fig:phasecurve_comparison} shows the resulting reflected-light phase curves for a TRAPPIST-1b-like configuration at $\lambda=0.55\,\mu$m using the Apollo regolith analog from the BUG dataset.

The light-blue region indicates the uncertainty envelope obtained from single-geometry conditioning alone, while the green region shows the result after introducing grain-scattering priors. Although the additional priors reduce the allowed parameter space, substantial uncertainty remains in the predicted phase curves because the opposition-surge parameters remain poorly constrained.

The red envelope in Fig.~\ref{fig:phasecurve_comparison} shows the corresponding predictions obtained using the full multi-angular BUG constraints. In this case, the uncertainty collapses to a narrow region, demonstrating that direct angular measurements strongly constrain the backscattering behavior of the surface.

The effect is particularly pronounced at small phase angles ($\alpha\lesssim20^\circ$), where opposition effects contribute significantly to the reflected signal. Importantly, the phase-angle range relevant for secondary-eclipse ingress in a TRAPPIST-1b-like configuration ($\alpha\sim3^\circ$) lies precisely in the regime where the predictions are most sensitive to the poorly constrained opposition-surge parameters.

This experiment demonstrates that grain-scale optical measurements provide valuable constraints on microscopic scattering properties, but are insufficient to uniquely determine the macroscopic angular redistribution of reflected light from a regolith surface. Direct multi-angular BRDF measurements remain necessary to construct predictive reflected-light models of rocky exoplanets.

\begin{figure}[h!]
\centering
\includegraphics[width=\hsize]{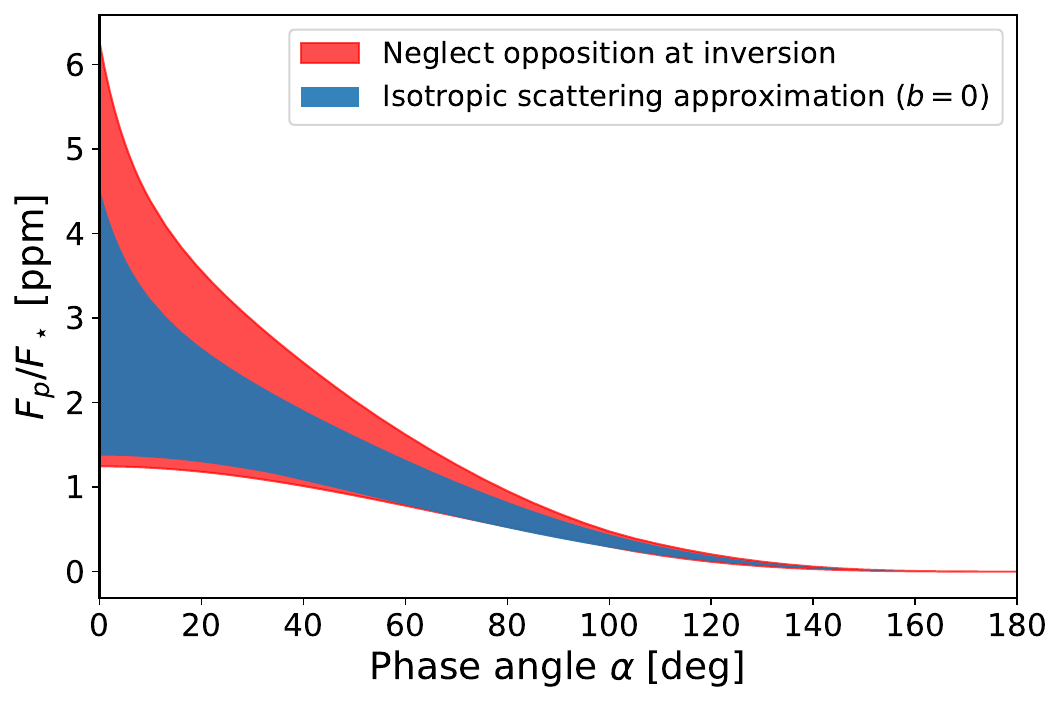}
\caption{
Disk-integrated reflected-light phase curves obtained under simplified photometric assumptions for the TRAPPIST-1b example considered in this work. The blue shaded region shows the uncertainty envelope obtained when the phase-function asymmetry parameter is fixed ($b=0$) while retaining the full photometric model. The red shaded region corresponds to the case where opposition effects are neglected during the inversion stage, allowing the single-scattering albedo to be uniquely inferred at the laboratory geometry and then propagated using the full model. In both cases, the solutions reproduce the same laboratory reflectance at the reference geometry but retain broad uncertainty when extrapolated to other phase angles. The two approximations produce broadly similar phase-curve morphologies, with the largest uncertainty occurring toward small phase angles where opposition effects strongly influence the reflected signal.
}
\label{fig:Approximations}
\end{figure}

\subsection{Impact of simplified photometric assumptions}

A common strategy in exoplanet modeling is to reduce the dimensionality of the photometric problem through simplified assumptions. Two approximations are frequently adopted: (i) fixing the phase-function asymmetry parameter, often assuming isotropic scattering ($b=0$), and (ii) neglecting opposition effects at moderate phase angles when interpreting laboratory measurements.

We first consider the impact of fixing the phase function. Although setting $b=0$ formally reduces the dimensionality of the parameter space, it does not eliminate the uncertainty in the predicted observables. Instead, the uncertainty shifts into the remaining free parameters, particularly the opposition-surge parameters $(B_0,h)$ and the single-scattering albedo $\omega$. In some wavelength regimes, the spread in the predicted phase curves can even increase relative to the fully unconstrained case.

We next examine a more restrictive approximation in which opposition effects are neglected at the laboratory measurement geometry ($g\sim30^\circ$), allowing the single-scattering albedo to be uniquely inferred from the measured reflectance. This inferred albedo is then propagated to other phase angles using the full photometric model.

The resulting phase curves are shown in Fig.~\ref{fig:Approximations}. The blue shaded region corresponds to the case where the phase-function asymmetry parameter is fixed ($b=0$), while the red shaded region corresponds to the case where opposition effects are neglected during the inversion stage.

The two approximations produce broadly similar phase-curve morphologies, but neither removes the large range of disk-integrated signals permitted by the single-geometry laboratory constraint. The uncertainty is largest toward small phase angles, where the reflected signal is particularly sensitive to the poorly constrained opposition behavior. Thus, the important result is not a strong morphological divergence between the two simplified prescriptions, but that imposing either assumption fails to uniquely determine the phase curve away from the laboratory measurement geometry.

Importantly, even at the orbital phase corresponding approximately to the laboratory measurement geometry, the disk-integrated signal does not collapse to a single value. This reflects the fact that the observable planetary flux arises from a distribution of local scattering geometries across the illuminated planetary disk. Matching the reflectance at a single laboratory geometry therefore does not uniquely constrain the disk-integrated observable.

These results demonstrate that simplified photometric assumptions do not resolve the underlying limitation introduced by incomplete angular laboratory constraints. Even when the inversion problem becomes formally well-defined, substantial uncertainty and systematic bias remain when extrapolating to the geometries relevant for exoplanet observations.

\subsection{Observational consequences for reflected-light missions}

The uncertainty associated with incomplete angular laboratory constraints can be quantified directly at the level of observable exoplanet signals. Rather than remaining confined to photometric parameter space, the uncertainty propagates into reflected-light phase curves, geometric albedos, and planet--star contrast ratios relevant for current and future high-contrast observations.

A particularly useful diagnostic is the geometric albedo $A_g$, which controls the reflected brightness of a planet near opposition:

\begin{equation}
\frac{F_p}{F_\star}
=
A_g
\left(
\frac{R_p}{a}
\right)^2.
\end{equation}

Because $A_g$ depends sensitively on the angular redistribution of reflected light at small phase angles, it provides a compact measure of how uncertainty in the photometric scattering properties propagates into observable reflected-light signals.

We propagate the posterior samples obtained from the retrievals in Section~\ref{sec:degeneracy_breaking} into predictions of $A_g$ under three conditioning schemes:

\begin{enumerate}
\item single-geometry laboratory conditioning,
\item single-geometry conditioning supplemented by grain-optics priors,
\item full multi-angular laboratory conditioning.
\end{enumerate}

\begin{figure}[h!]
\centering
\includegraphics[width=\hsize]{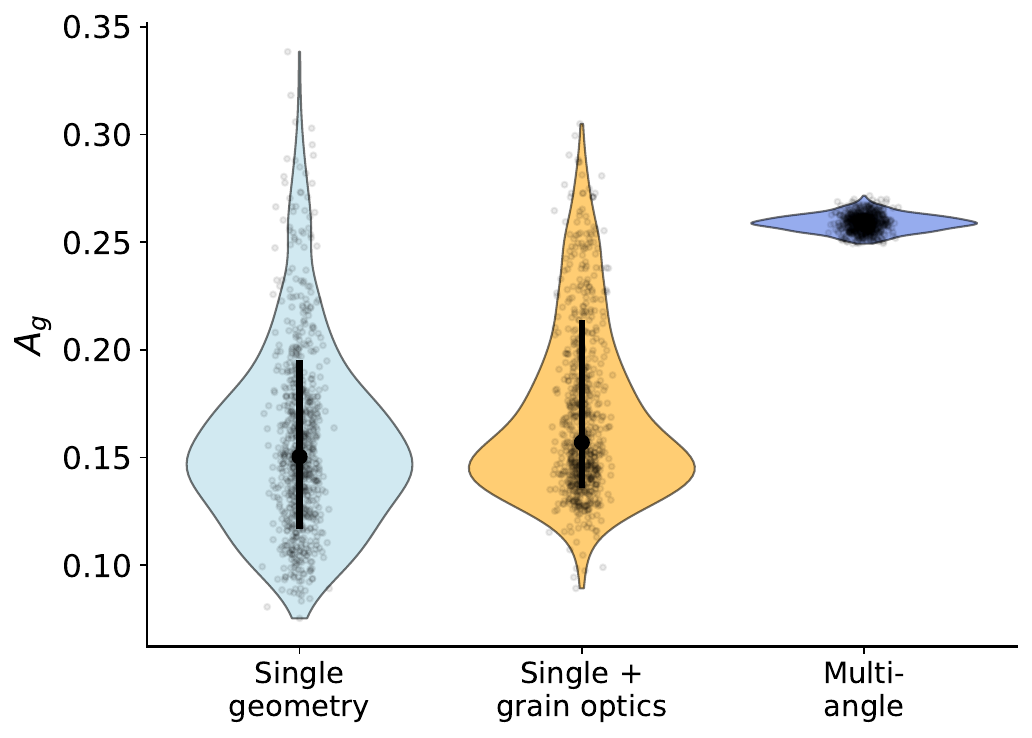}
\caption{
Posterior distributions of the geometric albedo $A_g$ predicted under the three conditioning schemes considered in this work. The violin distributions show the posterior geometric albedo $A_g$ for the single-geometry retrieval (left), the single-geometry retrieval supplemented by grain-scale optical constraints (center), and the full multi-angular retrieval (right). Black points and vertical bars indicate the posterior medians and 16th--84th percentile intervals, respectively. The individual posterior samples are shown as faint black points.
}
\label{fig:Ag_violin}
\end{figure}

The resulting posterior distributions are shown in Fig.~\ref{fig:Ag_violin}. When only a single laboratory geometry is used, the predicted geometric albedo spans a broad interval,
\[
A_g = 0.148^{+0.042}_{-0.034}.
\]

Introducing grain-scale optical priors only modestly reduces the spread,
\[
A_g = 0.155^{+0.057}_{-0.019},
\]
because the opposition-surge behavior remains poorly constrained.

In contrast, conditioning the inversion on the full multi-angular dataset collapses the posterior to a narrow interval,
\[
A_g = 0.259^{+0.004}_{-0.004},
\]
reducing the credible interval by approximately an order of magnitude.

The comparison additionally reveals a systematic shift in the predicted brightness. In this example, single-geometry conditioning systematically favors lower geometric albedo solutions relative to the multi-angular result. Thus, incomplete angular laboratory constraints may introduce not only large uncertainty envelopes, but also systematic biases in the predicted brightness of rocky exoplanets.

The resulting uncertainty propagates directly into observable planet--star contrast ratios relevant for future reflected-light missions. At small phase angles ($\alpha\lesssim10^\circ$), corresponding to secondary-eclipse geometries, the ensemble of photometric solutions consistent with a single laboratory geometry produces a broad range of predicted contrasts. For the example shown in Fig.~\ref{fig:phasecurve_comparison} at $\lambda=0.55\,\mu$m, the predicted reflected-light signal spans approximately $1.5$--$4.5$~ppm, corresponding to a fractional uncertainty $\delta_{\rm model}\sim0.5$.

In the direct-imaging regime ($\alpha\sim60^\circ$--$90^\circ$), the impact of the uncertainty depends on wavelength. In reflection-dominated bands, the uncertainty remains large, with $\delta_{\rm model}\sim0.5$--$0.6$. In contrast, in bands where thermal emission contributes significantly to the signal, the uncertainty decreases to $\delta_{\rm model}\sim0.1$--$0.2$.

Simplified photometric assumptions introduce additional systematic biases. Models assuming isotropic scattering and neglecting opposition effects underestimate the reflected flux at small phase angles by $\sim50$--$65\%$, while overestimating the signal at larger phase angles by more than a factor of two.

For Habitable Worlds Observatory (HWO)-like observations of Earth analogs in reflected light, the characteristic contrast scale is $C_p\sim10^{-10}$ \citep{Astro2020}. Adopting $\delta_{\rm model}\sim0.1$--$0.5$ yields

\begin{equation}
\Delta C_{\rm model}^{\rm HWO}
\sim
(1\text{--}5)\times10^{-11},
\end{equation}

comparable to the contrast-stability requirements discussed for future space-based coronagraphs \citep{Astro2020,Origins2023}.

For ELT-class facilities, where expected reflected-light contrasts are typically $C_p\sim10^{-9}$--$10^{-8}$ \citep{Hook2009,Kasper2021}, the corresponding uncertainty becomes

\begin{equation}
\Delta C_{\rm model}^{\rm ELT}
\sim
(1\text{--}5)\times10^{-10}
\end{equation}

to

\begin{equation}
\Delta C_{\rm model}^{\rm ELT}
\sim
(1\text{--}5)\times10^{-9}.
\end{equation}

These results demonstrate that the uncertainty associated with incomplete angular laboratory constraints is not merely a formal ambiguity in photometric parameter space. In reflected-light regimes, it propagates directly into uncertainties and systematic biases in planet--star contrast ratios comparable to the signal levels targeted by future high-contrast observations. Current laboratory datasets are therefore insufficient for uniquely predictive reflected-light modeling of rocky exoplanets without additional angular constraints on the surface scattering behavior.

\section{Discussion and conclusions}

The results presented in this work demonstrate that the current experimental foundation used to model rocky exoplanet surfaces contains an important structural limitation. Most laboratory spectral libraries provide reflectance measurements at only a single illumination-viewing geometry. While such measurements constrain the spectral reflectance of planetary materials, they do not uniquely determine the angular scattering behavior required to predict disk-integrated reflected-light observables.

Using the Hapke photometric framework, we showed that multiple photometric solutions can reproduce the same single-geometry laboratory reflectance spectrum while predicting substantially different reflected-light phase curves, geometric albedos, and planet--star contrast ratios. This uncertainty propagates directly into observable exoplanet signals and remains significant across reflection-dominated wavelength regimes relevant for future direct-imaging observations.

Using multi-angular laboratory measurements from the BUG dataset, we demonstrated that angular diversity strongly reduces the allowed photometric parameter space and collapses the uncertainty in predicted reflected-light observables. The dominant limitation is therefore not necessarily the photometric formalism itself, but the incomplete angular coverage of the laboratory datasets used to constrain it.

We additionally explored whether indirect approaches could compensate for the absence of multi-angular laboratory measurements. Grain-scale optical constraints reduce the allowed parameter space but do not uniquely constrain the macroscopic angular redistribution of reflected light from granular surfaces. Similarly, simplified photometric assumptions frequently adopted in exoplanet studies, such as isotropic scattering or neglecting opposition effects, introduce substantial systematic biases when extrapolated to observational geometries relevant for exoplanets.

The resulting uncertainty is observationally significant. In reflected-light regimes, incomplete angular constraints propagate into planet--star contrast uncertainties comparable to the signal levels targeted by future high-contrast facilities such as HWO and ELT-class instruments. In several cases, the systematic bias introduced by simplified photometric assumptions exceeds the intrinsic uncertainty associated with the photometric solutions themselves.

These results indicate that current laboratory reflectance datasets are insufficient for uniquely predictive reflected-light modeling of rocky exoplanets. Expanding laboratory BRDF measurements to include broad multi-angular coverage across a wider range of planetary analog materials, grain sizes, textures, and physical states is therefore an important next step. The present work is intended to motivate this experimental effort and to identify the angular information that is currently missing.

Once sufficiently broad multi-angular laboratory databases become available, these measurements could form the basis of Bayesian surface-retrieval frameworks in which composition, grain size, and other physical surface properties are inferred from reflected-light observations while accounting for their wavelength- and geometry-dependent scattering behavior. In this sense, laboratory BRDF databases could eventually play a role analogous to molecular opacity databases in atmospheric retrievals, providing the experimentally constrained inputs needed to connect observed spectra and phase-dependent brightness to the physical properties of a planetary surface.

More broadly, our results highlight that the challenge is not solely observational sensitivity, but also the physical completeness of the laboratory constraints used in exoplanet forward models. As reflected-light characterization enters the terrestrial exoplanet regime, realistic angular scattering measurements will become increasingly important for connecting observed planetary brightness to the physical properties of rocky surfaces.

\begin{acknowledgements}

\textit{Acknowledgments:}  L.G and F.J.P. acknowledge financial support from the Severo Ochoa grant CEX2021-001131-S funded by MCIN/AEI/10.13039/501100011033 and 
Ministerio de Ciencia e Innovación through the project PID2022-137241NB-C43.

\end{acknowledgements}

\bibliography{sample702}{}
\bibliographystyle{aasjournalv7.1}

\end{document}